# Instrumented shoulder functional assessment using inertial measurement units for frozen shoulder

Ting-Yang Lu
Department of Biomedical Engineering
National Yang Ming Chiao Tung University
Taipei 112, Taiwan
tylu.y@nycu.edu.tw

Kai-Chun Liu
Research Center for Information Technology Innovation
Academia Sinica
Taipei 115, Taiwan
t22302856@citi.sinica.edu.tw

Chia-Yeh Hsieh
Department of Biomedical Engineering
National Yang Ming Chiao Tung University
Taipei 112, Taiwan
kerrhsieh@nycu.edu.tw

Chih-Ya Chang
Department of Physical Medicine and Rehabilitation
Tri-Service General Hospital
Taipei 114, Taiwan
gradesboy@gm.ym.edu.tw

Yu Tsao
Research Center for Information Technology Innovation
Academia Sinica
Taipei 115, Taiwan
yu.tsao@citi.sinica.edu.tw

Chia-Tai Chan
Department of Biomedical Engineering
National Yang Ming Chiao Tung University
Taipei 112, Taiwan
ctchan@nycu.edu.tw

*Abstract*—Frozen shoulder (FS) is a shoulder condition that leads to pain and loss of shoulder range of motion. FS patients have difficulties in independently performing daily activities. Inertial measurement units (IMUs) have been developed to objectively measure upper limb range of motion (ROM) and shoulder function. In this work, we propose an IMU-based shoulder functional task assessment with kinematic parameters (e.g., smoothness, power, speed, and duration) in FS patients and analyze the functional performance on complete shoulder tasks and subtasks. Twenty FS patients and twenty healthy subjects were recruited in this study. Five shoulder functional tasks are performed by participants, such as washing hair (WH), washing upper back (WUB), washing lower back (WLB), placing an object on a high shelf (POH), and removing an object from back pocket (ROP). The results demonstrate that the used smoothness features can reflect the differences of movement fluency between FS patients and healthy controls ($p < 0.05$ and effect size $> 0.8$). Moreover, features of subtasks provided subtle information related to clinical conditions that have not been revealed in features of a complete task, especially the defined subtask 1 and 2 of each task.

I. INTRODUCTION

Adhesive capsulitis, also called frozen shoulder (FS), is a common shoulder condition characterized by painful and gradual loss of passive and active shoulder motion. The prevalence of FS is estimated at 2 % to 5% of general population. FS commonly occurred in woman between 40 and 60 years of age [1]. Patients with FS are commonly restricted to movements involved in forward flexion and external rotation [2]. Shoulder conditions can significantly affect their ability to carry out daily activities and work.

Accurate shoulder function assessment is essential to evaluate the severity of illness and effectiveness of treatments in clinical practice. The measures in routine clinical settings involve measuring range of motion (ROM) [3] and patient-reported questionnaires [4]. However, there are several challenges limiting these typical methods. For example, the measurement of ROM mainly relies on manual operation, which may be prone to errors and biases during the assessment process. Moreover, patient-reported questionnaires have controversial issues due to variety of languages and culture, interpretations of tester and subject and content validity [5].

In recent years, inertial-measurement-unit based (IMU-based) human motion analysis systems have been progressively implemented in clinical settings to estimate motion characteristics and to evaluate movement quality [6]. These systems have the advantages of portability, user-friendliness and lower cost. Several well-known studies have developed IMU-based systems to objectively measure upper limb range of motion for clinical assessment [7, 8]. However, evaluating ROM is insufficient to present the capacity to carrying out a functional task in FS patients.

To assess the functional ability of the shoulder, a previous work have proposed IMU-based approach to extract useful kinematic features to evaluate shoulder function in FS patients [9]. Several power related features, such as P score and M score, quantified the kinematic differences between painful shoulder and healthy shoulder. Features of smoothness are also essential which reflect healthy and well-trained motor behavior [10]. Movement smoothness has been seen as an indicator of upper extremity motor recovery in stroke patients [11]. However, the smoothness of the shoulder tasks have not been investigated in FS patients.

To investigate more detailed information about the movement patterns, numerous IMU-based motion analysis systems have divided the complicated task into multiple subtasks for clinical analysis. Van Uem et al. [12] differentiated mild-to-moderate Parkinson's disease patients from controls using durations and kinematic parameters of sub-phases of instrumented-Timed-Up-and-Go (iTUG) test. Their experimental results revealed that sub-phases could provide more information of subtle movement deficits in PD patients. Nevertheless, most previous works only focused on the complete shoulder task or a single subtask for FS assessment [9, 13]. For example, the performance of a complete shoulder task "put hand to a back" can be divided into three subtasks, involving "lifting hands toward back", "holding behind back", and "putting hands back". Only the performance of complete task have been investigated [13].

In this study, we propose instrumented shoulder functional assessment using IMUs for FS. The aim of this study is to validate the properties of kinematic features to reflect the differences in shoulder function between FS patients and healthy subjects. Investigation of subtasks is hypothesized to obtain more information on functional performance implicit in the complete task.

## II. MATERIALS AND METHODS

### A. Participants

A total of 40 participants were recruited in this study. Twenty patients (mean age: 57.6±11.8, range: 36-79 years old, height: 162.3±7.4 cm, weight: 58.4±9.5 kg, 13 female, 8 right arm affected) with a diagnosis of FS suffered from shoulder pain and limited shoulder ROM. Diagnoses were made by a physiatrist at the rehabilitation department of Tri-service general hospital. The exclusion criteria is as follows: bilateral shoulder conditions, full or massive thickness tear of the rotator cuff on magnetic resonance imaging (MRI) or ultrasonography, secondary FS, and acute cervical radiculopathy [14]. Twenty healthy subjects (mean age: 24.6±3.8, range: 21-35 years old, height: 168.6±6.7 cm, weight: 68±15.3 kg, 10 female, 17 right arm dominant) without the history of shoulder condition were recruited for a comparison group. This study was approved by the institutional review board (TSGHIRB No.: A202005024) at the university hospital. All participants were provided informed consent and entirely voluntary for their participation.

### B. Assessment protocol

Two IMUs are placed on the wrist and the arm to acquire movement signal of functional shoulder tasks, as shown in Fig. 1. Each IMU (APDM Inc., Portland, USA) including a tri-axial accelerator (range: ±16 g, resolution: 14 bits) and a tri-axial gyroscope (range: ±2000 °/s, resolution: 16 bits) recorded data with the sampling rate of 128Hz.

During the assessment procedure, IMUs are attached on the painful side of the hand for the patient, and that on the dominant side of the hand for the healthy subject. Each participant is instructed to perform 5 shoulder tasks based on the Shoulder Pain and Disability Index (SPADI), including washing hair (WH), washing upper back (WUB), washing lower back (WLB), placing an object on a high shelf (POH), and removing an object from back pocket (ROP), as listed in TABLE I. These tasks are related to subtle movement dimensions that patients commonly be restricted, including forward flexion, external rotation. The initial position of each task is standing with feet at hip-width, and arms alongside the body. Participants are given oral expression of each task and presenting self-paced movement in their habitual style. The assessment is filmed with video for tasks labelling and carried out before the intervention therapy.

### C. Data pre-processing

In order to acquire more detailed information associate with shoulder pain and limited ROM, we manually label and segment the complete shoulder tasks and subtasks from consecutive recording data. Each complete shoulder task is divided into 3 subtasks. The first and the third subtasks are similar to reaching task [15], and the second subtask is the main activity, as listed in TABLE I. For each segment (involving complete shoulder task and subtasks), $a_x, a_y, a_z$ represent the x-, y-, z-axial acceleration ($a$) respectively, and so on, $\omega_x, \omega_y, \omega_z$ represent the x-, y-, z-axial angular velocity ($\omega$). An example of the signals performed by the subject is illustrated in Fig. 2. Then, $a_{Norm}, \omega_{Norm}$ denoted as Euclidean norm of tri-axial acceleration and angular velocity are calculated by equation (1) and (2). Both motion signals collected from the wrist and arm are used to for IMU feature extraction.

$$a_{Norm} = \sqrt{a_x^2 + a_y^2 + a_z^2} \quad (1)$$

$$\omega_{Norm} = \sqrt{\omega_x^2 + \omega_y^2 + \omega_z^2} \quad (2)$$

### D. IMU features

After the data pre-processing, various kinematic features are applied to the complete shoulder task and subtasks. There are two types of IMU features that are extracted for performance analysis.

The first one is smoothness. Movement smoothness is a quantity to measure movement non-intermittency. Several methods have been developed to quantitative movement smoothness in different aspects of movement characteristics to provide interpretation based on nature of movement. The illustration of smoothness features are as follows:

a) **Number of mean crossing points (NMCP-A):** NMCP-A is a quantity measurement of the change of movement direction during performing the task. The calculation of NMCP-A is based on $a_{Norm}$.

b) **Number of peaks in acceleration (NP-A):** The number of peaks increased represents less smoothness during performing the task. The calculation of NP-A is based on $a_{Norm}$.

c) **Spectral arc length (SPARC):** SPARC is the arc length of the Fourier transform of the velocity profile of movement segment [16]. A movement composed of multiple sub-movements would result in a longer arc length indicated less smoothness. The calculation of SPARC is based on $\omega_{Norm}$.

TABLE I. SHOULDER TASKS DESCRIPTION

| Task | Subtask | Explanation |
|---|---|---|
| Washing hair (WH) | 1 | Lift up hands toward the top of the head |
| | 2 | Wash hair for few seconds |
| | 3 | Put down hands and return to the initial position |
| Washing upper back (WUB) | 1 | Lift up hand toward neck |
| | 2 | Washing the upper back for few seconds, including shoulders on both sides and the neck |
| | 3 | Put down the hand and return to the initial position |
| Washing lower back (WLB) | 1 | Rotate the hand toward the back |
| | 2 | Wash the lower back for few seconds, involving the area between the shoulder blade and the waist |
| | 3 | Put down the hand, return to the start position |
| Placing an object on a high shelf (POH) | 1 | Hold a smartphone using the painful/dominant hand in the initial position, then lift hand to the height approximately above the head |
| | 2 | Holding the phone in the air for few seconds |
| | 3 | Put down the hand and return to the initial position |
| Removing an object from back pocket (ROP) | 1 | Hold a smartphone the painful/dominant hand in the initial position, then rotate the hand toward the back pocket |
| | 2 | Put the phone into the pocket then take out of it |
| | 3 | Put down the hand and return to the initial position |

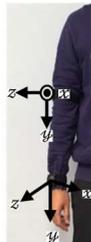

Fig. 1. IMUs placement and axes on the right side of the hand

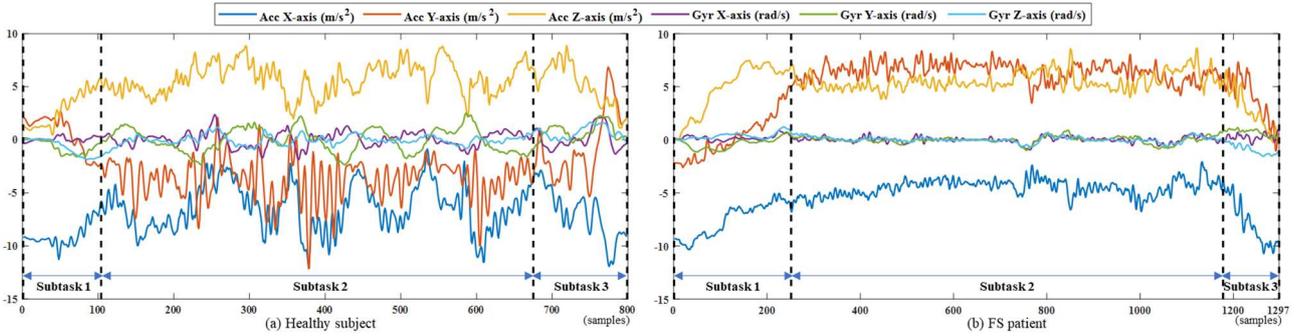

Fig. 2. An example of the IMU signal of WLB performed by (a) a healthy subject (b) an FS patient, where a complete task is divided into 3 subtasks

d) **Log dimensionless jerk in acceleration (LDLJ-A)**: LDLJ-A is the natural log of the sum of the squared jerk multiplied with the task duration and divided by the squared peak acceleration [17]. The calculation of LDLJ-A is based on $a_{Norm}$.

The second type is to extract movement characteristics related to power and speed during the performance of shoulder tasks and subtasks. The selected 3 IMU features are as follows:

e) **Range of Angular Velocity (RAV):** RAV is calculated by the mean of the range in tri-axial angular velocity, reflected the speed change during the task [9].
f) **Power Index (PI):** PI is calculated by the product of the range of acceleration and the range of angular velocity related to power control of the estimated part of the body [9].
g) **Duration:** Movement duration is defined as the period where subject performs the task by manual labeling the recorded video.

*E. Statistical analysis*

The independent t-test is used to show if there are significant differences between two groups for each feature of complete shoulder tasks and subtasks. With $p$ values $< 0.05$ are considered statistically significant. The responsiveness of the two groups was calculated using Cohen's d effect size with a 95% confidence interval was used to determine if effect sizes large ($d \geq 0.8$) [18]. Statistical analyses were performed using IBM SPSS® Statistics version 24 (IBM, Armonk, USA).

### III. RESULTS AND DISCUSSION

All participants are able to accomplish the shoulder functional assessment. There are total 200 complete shoulder tasks and 600 subtasks collected from the FS patient and the healthy groups. IMU features derived from each segment of two IMU placements are used to analyze differences between two groups by independent *t*-test and effect size (Cohen's *d*). The experimental results for each shoulder task are shown in TABLE II -TABLE VI.

The differences between FS patients and healthy subjects are statistically significant, with four smoothness features for all tasks. However, the significant impact of the metrics is not in all subtasks. The smoothness features mainly show the significant differences in subtask 1 of all tasks and subtask 2 of WUB and POH, while that exhibit only one smoothness feature has the significant impact in subtask 3. It revealed that the smoothness features are sensitive to movements of flexion & extension (subtask 1) and cleaning & holding (subtask 2). The results demonstrate that a single smoothness feature is insufficient to present the differences in movement qualities between groups. The calculated smoothness features play different roles in each shoulder task. There is a requirement to design or apply a more general feature for smoothness assessment in shoulder tasks. Furthermore, the smoothness features of ROP can distinguish differences between patient and healthy groups in subtask 1, but not in the complete task.

TABLE II. THE P VALUE FOR WH OF IMU FEATURES IN TWO PLACEMENTS

| Parameter | Placement | Washing hair | | | |
|---|---|---|---|---|---|
| | | Complete Task | Subtask 1 | Subtask 2 | Subtask 3 |
| NMCP-A | Wrist | 0.088 | 0.077 | 0.205 | 0.135 |
| | Arm | 0.045 | 0.104 | 0.135 | 0.046 |
| NP-A | Wrist | 0.005* | 0.004* | 0.024 | 0.01 |
| | Arm | 0.028 | 0.001* | 0.224 | 0.014 |
| LDLJ-A | Wrist | 0.947 | 0.003* | 0.382 | 0.395 |
| | Arm | 0.528 | 0.002* | 0.546 | 0.385 |
| SPARC | Wrist | 0.083 | 0.122 | 0.608 | 0.885 |
| | Arm | 0.186 | 0.367 | 0.607 | 0.617 |
| RAV | Wrist | 0.005 | 0.031 | 0.008 | 0.035 |
| | Arm | 0.002* | 0.005* | 0.008 | 0.014 |
| PI | Wrist | 0.002* | 0.068 | 0.001* | 0.066 |
| | Arm | 0.005 | 0.045 | 0.004 | 0.078 |
| Duration | N/A | 0.017 | 0.005* | 0.1 | 0.012 |

*: $p < 0.05$ and Cohen's $d > 0.8$

TABLE III. THE P VALUE FOR WUB OF IMU FEATURES IN TWO PLACEMENTS

| Parameter | Placement | Washing upper back | | | |
|---|---|---|---|---|---|
| | | Complete Task | Subtask 1 | Subtask 2 | Subtask 3 |
| NMCP-A | Wrist | 0.048 | 0.073 | 0.06 | 0.481 |
| | Arm | 0.301 | 0.063 | 0.424 | 0.678 |
| NP-A | Wrist | 0.005* | 0.017* | 0.016 | 0.03 |
| | Arm | 0.17 | 0.087 | 0.287 | 0.304 |
| LDLJ-A | Wrist | 0.784 | 0.038 | 0.014* | 0.459 |
| | Arm | 0.407 | 0.053 | 0.496 | 0.549 |
| SPARC | Wrist | 0.002* | 0.211 | 0.025* | 0.24 |
| | Arm | 0.017 | 0.055 | 0.104 | 0.122 |
| RAV | Wrist | 0.002* | 0.077 | <0.001* | 0.004* |
| | Arm | <0.001* | <0.001* | 0.001* | 0.001* |
| PI | Wrist | 0.001* | 0.076 | 0.001* | 0.001* |
| | Arm | 0.001* | <0.001* | 0.006 | 0.004* |
| Duration | N/A | 0.007* | 0.048 | 0.016* | 0.069 |

*: $p < 0.05$ and Cohen's $d > 0.8$

TABLE IV. THE P VALUE FOR WLB OF IMU FEATURES IN TWO PLACEMENTS

| Parameter | Placement | Washing lower back | | | |
|---|---|---|---|---|---|
| | | Complete Task | Subtask 1 | Subtask 2 | Subtask 3 |
| NMCP-A | Wrist | 0.201 | 0.015 | 0.419 | 0.452 |
| | Arm | 0.015* | 0.087 | 0.057 | 0.008* |
| NP-A | Wrist | 0.342 | 0.286 | 0.47 | 0.252 |
| | Arm | 0.154 | 0.341 | 0.288 | 0.022 |
| LDLJ-A | Wrist | 0.13 | 0.012* | 0.878 | 0.249 |
| | Arm | 0.156 | 0.006* | 0.73 | 0.052 |
| SPARC | Wrist | 0.416 | 0.001* | 0.401 | 0.17 |
| | Arm | 0.001* | 0.016 | 0.142 | 0.033 |
| RAV | Wrist | <0.001* | 0.001* | <0.001* | <0.001* |
| | Arm | <0.001* | <0.001* | <0.001* | <0.001* |
| PI | Wrist | <0.001* | <0.001* | <0.001* | <0.001* |
| | Arm | <0.001* | <0.001* | <0.001* | 0.001* |
| Duration | N/A | 0.114 | 0.034 | 0.251 | 0.139 |

*: $p < 0.05$ and Cohen's $d > 0.8$

TABLE V. THE P VALUE FOR POH OF IMU FEATURES IN TWO PLACEMENTS

| Parameter | Placement | Placing an object on a high shelf | | | |
|---|---|---|---|---|---|
| | | Complete Task | Subtask 1 | Subtask 2 | Subtask 3 |
| NMCP-A | Wrist | 0.049 | 0.055 | 0.001* | 0.257 |
| | Arm | 0.051 | 0.007* | 0.005* | 0.93 |
| NP-A | Wrist | 0.001* | 0.007* | 0.013* | 0.169 |
| | Arm | 0.141 | 0.06 | 0.576 | 0.843 |
| LDLJ-A | Wrist | 0.4 | 0.062 | 0.009 | 0.614 |
| | Arm | 0.309 | 0.06 | 0.064 | 0.885 |
| SPARC | Wrist | 0.872 | 0.627 | 0.008* | 0.66 |
| | Arm | 0.388 | 0.211 | 0.001* | 0.459 |
| RAV | Wrist | <0.001* | <0.001* | 0.034 | <0.001* |
| | Arm | <0.001* | <0.001* | 0.015 | <0.001 |
| PI | Wrist | <0.001* | <0.001* | 0.02 | <0.001* |
| | Arm | <0.001* | <0.001* | 0.046 | 0.002 |
| Duration | N/A | <0.001* | 0.004* | 0.01* | 0.206 |

*: $p < 0.05$ and Cohen's $d > 0.8$

TABLE VI. THE P VALUE FOR ROP OF IMU FEATURES IN TWO PLACEMENTS

| Parameter | Placement | Removing an object from back pocket | | | |
|---|---|---|---|---|---|
| | | Complete Task | Subtask 1 | Subtask 2 | Subtask 3 |
| NMCP-A | Wrist | <0.001 | 0.086 | 0.367 | 0.487 |
| | Arm | <0.001 | 0.02* | 0.861 | 0.733 |
| NP-A | Wrist | <0.001 | 0.025* | 0.162 | 0.991 |
| | Arm | 0.002 | 0.048 | 0.609 | 0.732 |
| LDLJ-A | Wrist | 0.002 | 0.197 | 0.192 | 0.649 |
| | Arm | 0.001 | 0.072 | 0.282 | 0.881 |
| SPARC | Wrist | 0.005 | 0.008* | 0.084 | 0.502 |
| | Arm | 0.006 | 0.003* | 0.065 | 0.518 |
| RAV | Wrist | <0.001 | 0.064 | 0.087 | 0.016 |
| | Arm | <0.001 | 0.085 | 0.025 | 0.006 |
| PI | Wrist | <0.001 | 0.152 | 0.056 | 0.03 |
| | Arm | <0.001* | 0.064 | 0.021 | 0.005* |
| Duration | N/A | <0.001 | 0.011* | 0.419 | 0.655 |

*: $p < 0.05$ and Cohen's $d > 0.8$

This is because most FS patients have more difficulties in rotating their hands to the back (subtask 1) instead of the whole task, which provides complementary movement information during the clinical assessment. The features extracted from the complete task may decrease the group discrepancy of the smoothness characteristics.

The features of RAV and PI derived from most sensor placements and subtasks have significant differences in WUB, WLB ,and POH while these features are less significant in WH and ROP. Such results show that WUB, WLB ,and POH are the more difficult shoulder tasks for FS patients. The revealed results are similar to previous works [9, 13]. Although shoulder tasks are made self-paced, the results demonstrate that feature of duration in several subtasks had a significant impact. The correlation between duration and movement pace control will be explored for further analysis. The calculated smoothness features are clinically meaningful and can reflect the differences in movement fluency between FS patients and healthy subjects. Moreover, we show that IMU features extracted from subtasks can provide subtle information to complement the complete task, especially subtask 1 and 2.

## IV. CONCLUSION

An accurate and reliable method to assess functional disorder for patients with FS is essential in clinical practice. This study proposed IMU-based shoulder functional assessment using various kinematic features to quantify differences of movement characteristics between FS patients and healthy subjects, which also could be used to differentiate. An analysis of shoulder subtasks can provide more function information for function evaluation. The proposed approach has the potential to provide more useful and elaborate measurement for FS assessment in clinical environments. There are several issues in the proposed functional assessment required to be further analyzed, such as the severity of FS, variety of clinical characteristics, and age differences. In future work, more subjects will be recruited to validate the reliability and generality of the proposed approach. A variety of shoulder-related daily activities (e.g., combing hair) will be involved to support clinical analysis. Furthermore, we plan to develop an automatic FS assessment system without manual operation, including shoulder task identification, subtask segmentation, and severity assessment.

ACKNOWLEDGMENT

This work was supported in part by grants from the Ministry of Science and Technology (MOST109-2221-E-010-005).

REFERENCES

[1] J. Ramirez, "Adhesive capsulitis: Diagnosis and management," American family physician, vol. 99, no. 5, pp. 297-300, 2019.
[2] R. Dias, S. Cutts, and S. Massoud, "Frozen shoulder," Bmj, vol. 331, no. 7530, pp. 1453-1456, 2005.
[3] K. Hayes, J. R. Walton, Z. L. Szomor, and G. A. Murrell, "Reliability of five methods for assessing shoulder range of motion," Australian Journal of Physiotherapy, vol. 47, no. 4, pp. 289-294, 2001.
[4] K. E. Roach, E. Budiman‐Mak, N. Songsiridej, and Y. Lertratanakul, "Development of a shoulder pain and disability index," Arthritis & Rheumatism: Official Journal of the American College of Rheumatology, vol. 4, no. 4, pp. 143-149, 1991.
[5] A. A. Ragab, "Validity of self-assessment outcome questionnaires: patient-physician discrepancy in outcome interpretation," Biomedical sciences instrumentation, vol. 39, pp. 579-584, 2003.
[6] I. H. Lopez-Nava and A. Muñoz-Meléndez, "Wearable inertial sensors for human motion analysis: A review," IEEE Sensors Journal, vol. 16, no. 22, pp. 7821-7834, 2016.
[7] C. P. Walmsley, S. A. Williams, T. Grisbrook, C. Elliott, C. Imms, and A. Campbell, "Measurement of upper limb range of motion using wearable sensors: a systematic review," Sports medicine-open, vol. 4, no. 1, pp. 1-22, 2018.
[8] M. Rigoni et al., "Assessment of Shoulder Range of Motion Using a Wireless Inertial Motion Capture Device—A Validation Study," Sensors, vol. 19, no. 8, p. 1781, 2019.
[9] B. Coley et al., "Outcome evaluation in shoulder surgery using 3D kinematics sensors," Gait & posture, vol. 25, no. 4, pp. 523-532, 2007.
[10] T. J. Sejnowski, "Making smooth moves," Nature, vol. 394, no. 6695, pp. 725-726, 1998.
[11] B. Rohrer et al., "Movement smoothness changes during stroke recovery," Journal of neuroscience, vol. 22, no. 18, pp. 8297-8304, 2002.
[12] J. M. Van Uem et al., "Quantitative timed-up-and-go parameters in relation to cognitive parameters and health-related quality of life in mild-to-moderate Parkinson's disease," PLoS One, vol. 11, no. 4, p. e0151997, 2016.
[13] C. Pichonnaz et al., "Heightened clinical utility of smartphone versus body-worn inertial system for shoulder function BB score," PLoS one, vol. 12, no. 3, p. e0174365, 2017.
[14] C.-Y. Chang et al., "Automatic Functional Shoulder Task Identification and Sub-Task Segmentation Using Wearable Inertial Measurement Units for Frozen Shoulder Assessment," Sensors, vol. 21, no. 1, p. 106, 2021.
[15] C. J. Newman et al., "Measuring upper limb function in children with hemiparesis with 3D inertial sensors," Child's Nervous System, vol. 33, no. 12, pp. 2159-2168, 2017.
[16] S. Balasubramanian, A. Melendez-Calderon, and E. Burdet, "A robust and sensitive metric for quantifying movement smoothness," IEEE transactions on biomedical engineering, vol. 59, no. 8, pp. 2126-2136, 2011.
[17] A. Melendez-Calderon, C. Shirota, and S. Balasubramanian, "Estimating Movement Smoothness from Inertial Measurement Units," bioRxiv, 2020.
[18] J. Cohen, Statistical power analysis for the behavioral sciences. Academic press, 2013.